\documentclass[conference]{IEEEtran}

\usepackage{cite}
\usepackage{amsmath,amssymb,amsfonts}
\usepackage{graphicx}
\usepackage{textcomp}
\usepackage{xcolor}
\usepackage{booktabs}
\usepackage[caption=false,font=footnotesize]{subfig}
\usepackage{xurl}

\newcommand{\bX}{\mathbf{X}}

\begin{document}

\title{SFPF: Spatio-Frequency Polarization Fingerprint for Anomalous Wireless Device Detection}
\author{Xiaoxuan Huang, Jinlong Xu, Daoyuan Shen, Meng Zhang, Dong Wei}
\maketitle

\begin{abstract}
Periodic inspection of deployed wireless devices is necessary because unauthorized hardware replacement may preserve communication functions, credentials, and logical identity, making anomalous devices difficult to detect. Such inspections are conducted under controlled measurement conditions to verify that each device remains consistent with its enrolled hardware state. Conventional radio-frequency fingerprint (RFF) may provide insufficient separation when replacement hardware closely resembles legitimate hardware, while a polarization fingerprint (PF) constructed at one observation direction may miss spatially nonuniform polarization changes. This paper proposes the spatio-frequency polarization fingerprint (SFPF), which jointly represents complex polarization responses over multiple frequencies and observation directions; conventional PF is its fixed-direction slice. We derive SFPF formation from hardware-dependent modal excitation, directional far-field radiation, and polarization projection. A first-order sensitivity analysis shows that the response to the same hardware change varies with both frequency and direction, motivating joint spatio-frequency acquisition. Electromagnetic simulations confirm the nonuniform spatio-frequency sensitivity and show that, under the same observation budget, SFPF improves normalized distance, Fisher score, and the inter-/intra-class ratio over PF by 17.7\%, 45.8\%, and 11.3\%, respectively. Experiments show that SFPF consistently outperforms RFF and PF over 0--20~dB. At 15--20~dB, SFPF achieves anomalous-device F1 scores of 87.3--90.4\% and AUROC values of 85.4--95.5\%.
\end{abstract}


\section{Introduction}
Wireless devices deployed over extended periods face the risk of unauthorized changes to their internal hardware. An adversary may replace one or more hardware modules in a wireless device with unauthorized, counterfeit, or maliciously modified alternatives. Counterfeit networking equipment found in practice can closely imitate a genuine product while using a different internal implementation~\cite{withsecure2020fakecisco}. Such hardware may enable unauthorized network access, traffic interception, sensitive-information exfiltration, or persistent service disruption. More importantly, a replaced device may retain its communication functions, software configuration, credentials, and apparent identity. It can therefore continue to pass credential-based authentication and software integrity verification even though its physical hardware has changed. Conventional hardware-integrity inspection often requires taking the device out of service and transferring it to a dedicated platform for disassembly, internal imaging, or probe-based measurements, which may disrupt normal operation or damage the device~\cite{tehranipoor2010survey,zhou2021backside}. Detecting this threat requires periodic, non-intrusive inspection of wireless devices to determine whether each device still conforms to its enrolled legitimate hardware state.

Physical-layer detection based on radiated wireless signals provides a non-contact way to detect hardware anomalies. Radio-frequency fingerprint (RFF) extracts hardware-dependent imperfections using transient and steady-state signatures such as carrier-frequency offset and I/Q imbalance, as well as raw I/Q samples, spectra, time--frequency features, and learned representations~\cite{soltanieh2020review,peng2023supervised,nhem2025explainable}. In a stealthy replacement attack, however, the adversary attempts to make the unauthorized module as similar as possible to the original hardware, leaving only subtle differences between legitimate and anomalous devices. Conventional RFF relies primarily on time- and frequency-domain signal characteristics, which provide insufficient separability for such subtle hardware anomalies~\cite{wang2025device}.

In addition to time- and frequency-domain characteristics, wireless signals also contain polarization-domain characteristics. These polarization characteristics are affected by the hardware characteristics of the wireless device and form a device-dependent polarization fingerprint (PF)~\cite{xu2022polarization,xu2022specific}. PF exhibits both frequency-dependent and spatially dependent characteristics. However, existing PF methods construct each fingerprint at a single observation direction and do not fully exploit its spatial characteristics. When the polarization difference between a legitimate device and an anomalous device is weak at the selected direction, fixed-direction PF may omit useful anomaly information and provide limited separability.

To address this limitation, we propose the \emph{spatio-frequency polarization fingerprint} (SFPF) for periodic, non-intrusive hardware-anomaly inspection of wireless devices. SFPF jointly organizes complex polarization responses over multiple frequencies and observation directions, while a conventional PF is its fixed-direction slice. We derive its formation from modal excitation and directional far-field projection, and analyze the sensitivity of SFPF to hardware perturbations. A detection network (ADNet) then encodes each directional observation, fuses its direction coordinate, and aggregates the resulting feature set to identify enrolled devices and reject anomalous devices. Electromagnetic simulations and prototype experiments demonstrate that SFPF provides stronger separability between legitimate and anomalous devices than RFF and PF.

The main contributions are as follows:
\begin{itemize}
    \item We introduce SFPF as a joint frequency--direction polarization representation and establish conventional PF as its fixed-direction slice, thereby extending PF from a single observation direction to the joint spatio-frequency domain.
    \item We derive the SFPF formation mechanism from modal excitation, far-field radiation, and hardware-perturbation sensitivity. The analysis shows why hardware-induced polarization differences vary across frequency and observation direction, and why sampling multiple reliable directions reduces the risk of missing a hardware change.
    \item Electromagnetic simulations show that hardware changes produce different SFPF responses across frequencies and observation directions, and that joint spatio-frequency sampling improves separability. Prototype experiments further demonstrate that SFPF achieves better anomalous-device detection performance than RFF and PF.
\end{itemize}

\section{Related Work}
\textbf{Hardware-integrity sensing.} Existing methods detect counterfeit components and malicious modifications through physical inspection, logic testing, or side-channel measurements such as timing, power, and near-field electromagnetic emissions~\cite{tehranipoor2010survey,zhou2021backside}. Physical inspection relies on imaging, microscopy, or backside access and often requires opening the enclosure or accessing internal components~\cite{tehranipoor2010survey,zhou2021backside}. Logic testing and timing- or power-based methods generally require wired access to chip pins, supply lines, or test interfaces, together with carefully designed stimuli~\cite{tehranipoor2010survey}. Near-field electromagnetic methods can avoid direct electrical contact~\cite{agrawal2007trojan,huang2014electromagnetic,lee2024robust}, but still require probes close to the chip or circuit board and may require removing the enclosure or shielding. Repeating these operations for periodic inspection of deployed wireless devices is time-consuming and risks disturbing or damaging the device. This limitation motivates non-contact approaches that assess hardware integrity directly from the device's intended wireless transmissions.

\textbf{Radio-frequency fingerprint.} RFF extracts hardware-dependent characteristics from wireless signals, thereby enabling non-contact anomaly detection without opening the device or accessing its internal test interfaces~\cite{soltanieh2020review,peng2023supervised}. Existing methods use turn-on transients, steady-state impairments, spectrograms, and learned representations~\cite{soltanieh2020review,peng2023supervised,nhem2025explainable}. In a stealthy replacement attack, the claimed logical identity may remain valid while the unauthorized hardware closely imitates the original hardware. The resulting difference between legitimate and anomalous states can be small, reducing the separability of conventional RFF representations constructed primarily from time-, frequency-domain signal characteristics~\cite{soltanieh2020review}.

\textbf{Polarization fingerprint.} PF provides a non-contact representation for hardware-anomaly detection because hardware changes in a wireless device alter the polarization state of its signal~\cite{xu2022polarization,xu2023polarization}. Existing studies have established the formation mechanism of PF and identified its frequency-dependent and spatially dependent characteristics; PF has also been investigated for device identification, physical-layer authentication, spatial characterization, and artificial fingerprint injection~\cite{xu2022polarization,xu2022specific,xu2023polarization,wang2025device}. Nevertheless, each PF sample is still constructed at a single observation direction~\cite{xu2022polarization,xu2022specific,xu2023polarization}. Existing PF representations do not jointly model polarization variations across frequencies and observation directions, which limits their ability to expose subtle hardware anomalies.

In summary, existing hardware-integrity methods are often intrusive, RFF may provide insufficient separation for subtle replacement anomalies, and existing PF methods do not fully exploit the spatial characteristics of polarization, potentially discarding discriminative information. SFPF addresses this limitation by incorporating observation direction as a physical dimension and jointly modeling polarization variations across frequency and space. It therefore preserves more discriminative information about hardware changes for anomalous-device detection.

\section{SFPF Formation and Construction}
\label{sec:sfpf}

\subsection{Formation Mechanism}
\label{subsec:formation}

We use a single-radiating-element circularly polarized antenna to expose the principal physical mechanism underlying the spatio-frequency polarization fingerprint (SFPF). Its radiation is mainly dominated by two orthogonal near-degenerate characteristic modes. According to characteristic mode theory~\cite{harrington1971theory,chen2015characteristic}, the surface current can be approximated as
\begin{equation}
\mathbf{J}(\mathbf r',f)
=
a_1(f)\mathbf{J}_1(\mathbf r')
+
a_2(f)\mathbf{J}_2(\mathbf r'),
\label{eq:current}
\end{equation}
where $\mathbf{J}_n(\mathbf r')$ denotes the current distribution of the $n$th mode and $a_n(f)$ is its frequency-dependent complex excitation coefficient.

Around resonance, the modal coefficient can be approximated as
\begin{equation}
a_n(f)
=
\frac{V_n}
{1+jQ_n\left(f/f_n-f_n/f\right)},
\qquad n=1,2,
\label{eq:modal_coeff}
\end{equation}
where $V_n$, $Q_n$, and $f_n$ denote the feed-dependent excitation, quality factor, and resonant frequency, respectively. Hardware variations can perturb these quantities and thereby alter the relative modal amplitude and phase over frequency.

For an observation direction $(\theta,\phi)$, define
\begin{equation}
\hat{\mathbf r}(\theta,\phi)
=
\begin{bmatrix}
\sin\theta\cos\phi &
\sin\theta\sin\phi &
\cos\theta
\end{bmatrix}^{T}.
\label{eq:direction}
\end{equation}
Under the far-field approximation, the directional response of the $n$th mode and the total far-field vector response can be written as~\cite{balanis2016antenna}
\begin{equation}
\begin{aligned}
\mathbf F_n(k,\theta,\phi)
&=
\int_S
\mathbf J_n(\mathbf r')
e^{jk\hat{\mathbf r}(\theta,\phi)\cdot\mathbf r'}
\,dS',\\
\mathbf F(f,\theta,\phi)
&=
\sum_{n=1}^{2}
a_n(f)\mathbf F_n(k,\theta,\phi).
\end{aligned}
\label{eq:farfield}
\end{equation}
Frequency therefore changes the relative modal weights through $a_n(f)$, whereas observation direction changes the coherent addition of the distributed modal currents through $\mathbf F_n(k,\theta,\phi)$.

Because the far-field electric field is transverse to the propagation direction, it can be expressed as~\cite{balanis2016antenna}
\begin{equation}
\mathbf E(f,\theta,\phi)
=
C(r,f)
\left(
\mathbf I-
\hat{\mathbf r}\hat{\mathbf r}^{T}
\right)
\mathbf F(f,\theta,\phi),
\label{eq:electric_field}
\end{equation}
where $C(r,f)$ collects propagation terms common to the field components. With two orthogonal receive components $E_x$ and $E_y$, the complex polarization ratio is defined as~\cite{ludwig1973definition}
\begin{equation}
p(f,\theta,\phi)
=
\frac{E_x(f,\theta,\phi)}
     {E_y(f,\theta,\phi)}.
\label{eq:polarization_ratio}
\end{equation}

Equations~\eqref{eq:modal_coeff}--\eqref{eq:polarization_ratio} show that the polarization response inherently depends on both frequency and observation direction. Frequency determines how the hardware-dependent modes are excited, while direction affects both their far-field coherent addition and their projection onto the orthogonal receive components. Thus, changing the observation direction does not merely change the received signal strength; it changes the observed polarization state itself.

To characterize how a subtle hardware change appears in the spatio-frequency polarization response, let $\boldsymbol{\xi}$ collect effective hardware-dependent parameters governing modal excitation and radiation, and let $\Delta\boldsymbol{\xi}$ denote a small perturbation. A first-order expansion gives
\begin{equation}
\Delta p(f,\theta,\phi)
\approx
\nabla_{\boldsymbol{\xi}}p(f,\theta,\phi)^{T}
\Delta\boldsymbol{\xi}.
\label{eq:sensitivity}
\end{equation}
The sensitivity $\nabla_{\boldsymbol{\xi}}p(f,\theta,\phi)$ varies with both frequency and observation direction through \eqref{eq:modal_coeff}--\eqref{eq:electric_field}. Consequently, the same hardware perturbation can produce nonuniform polarization deviations over the spatio-frequency domain. The difference may be weak at one frequency or observation direction but pronounced at another. A fixed-direction PF may therefore miss discriminative hardware information when the selected direction is insensitive to the particular perturbation.

To describe how observable a hardware perturbation is at a given direction over the sampled frequency range, we define
\begin{equation}
D_{\mathrm{dir}}(\theta,\phi)
=
\sum_{l=1}^{N_f}
\left|
\Delta p(f_l,\theta,\phi)
\right|^2.
\label{eq:sfpf_observability}
\end{equation}
A larger $D_{\mathrm{dir}}(\theta,\phi)$ indicates that the hardware perturbation produces a more pronounced polarization difference at that direction, whereas a smaller value indicates a less informative direction. Since $D_{\mathrm{dir}}(\theta,\phi)$ is generally nonuniform over space, relying on a single observation direction can omit useful anomaly information.

This nonuniform sensitivity, rather than increased feature dimensionality alone, motivates joint spatio-frequency acquisition. By sampling multiple frequencies and observation directions, SFPF retains polarization responses from a broader set of spatial and spectral conditions and is therefore less likely to discard informative hardware-induced deviations.

\subsection{SFPF and Its Relation to PF}
\label{subsec:construction}

Let
\begin{equation*}
\mathcal F=\{f_l\}_{l=1}^{N_f}
\end{equation*}
denote the sampled frequency set, and let
\begin{equation*}
\mathcal R
=
\{(\theta_a,\phi_a)\}_{a=1}^{N_a}
\end{equation*}
denote the set of distinct physical observation directions. The SFPF of a device is defined as
\begin{equation}
\mathbf S
=
\left\{
p(f_l,\theta_a,\phi_a)
\right\}_{a=1,l=1}^{N_a,N_f}
\in
\mathbb C^{N_a\times N_f}.
\label{eq:sfpf}
\end{equation}
In practical acquisition, \(N_s\) complex polarization-ratio samples are retained for each direction--frequency pair, giving the sampled SFPF tensor
\(\widetilde{\mathbf S}\in\mathbb C^{N_a\times N_f\times N_s}\).

At a fixed observation direction $(\theta_0,\phi_0)$, a conventional polarization fingerprint (PF) is
\begin{equation}
\mathbf p_{\mathrm{PF}}(\theta_0,\phi_0)
=
\left\{
p(f_l,\theta_0,\phi_0)
\right\}_{l=1}^{N_f}
=
\mathbf S[a_0,:],
\label{eq:pf_slice}
\end{equation}
where $(\theta_{a_0},\phi_{a_0})=(\theta_0,\phi_0)$. Hence, conventional PF is a fixed-direction slice of SFPF.

PF preserves polarization variation over frequency at one observation direction, but it does not capture how the frequency-dependent polarization response varies across observation directions. In contrast, SFPF jointly preserves polarization variations over frequency and space. Because hardware-induced deviations can be spatially nonuniform, observations from multiple directions provide complementary information that may be absent from a fixed-direction PF.

\section{SFPF-Based Hardware Anomaly Detection}
\subsection{Threat Model}
\label{subsec:threat_model}

We consider periodic hardware-anomaly inspection of enrolled wireless devices. 
The inspection receiver, dual-polarized antenna, 
enrolled SFPFs, and trained detector are assumed to be trusted. 
Before deployment, each authorized device is measured in its legitimate hardware 
configuration and enrolled as a reference. An authorized hardware change requires 
re-enrollment; otherwise, any deviation from the enrolled hardware state is regarded 
as a hardware anomaly.

The adversary is assumed to have physical access to the device and may replace the 
antenna (A), RF front end (R), or digital/baseband module (D), with functionally compatible unauthorized
components. The compromised device may preserve its software configuration, credentials, 
protocol behavior, communication functions, and apparent logical identity, and may 
therefore continue to pass conventional upper-layer authentication. We assume that the adversary 
cannot compromise the trusted inspection chain, modify the enrolled fingerprints or 
detector parameters.

For the set of enrolled device identities $\mathcal{K}$, the detector maps an observed 
fingerprint $\mathbf{X}$ to
\begin{equation}
\mathcal{G}:\mathbf{X}\rightarrow
\mathcal{K}\cup\{\mathrm{anomalous}\}.
\label{eq:detection_mapping}
\end{equation}
The security objective is to accept devices that remain in their enrolled hardware states 
and reject devices whose hardware states deviate from their enrolled states. We evaluate all seven nonempty replacement combinations of the three 
modules: A, R, D, A+R, A+D, R+D, and A+R+D. 

\begin{figure}[t]
 \centering
 \includegraphics[width=0.99\columnwidth]{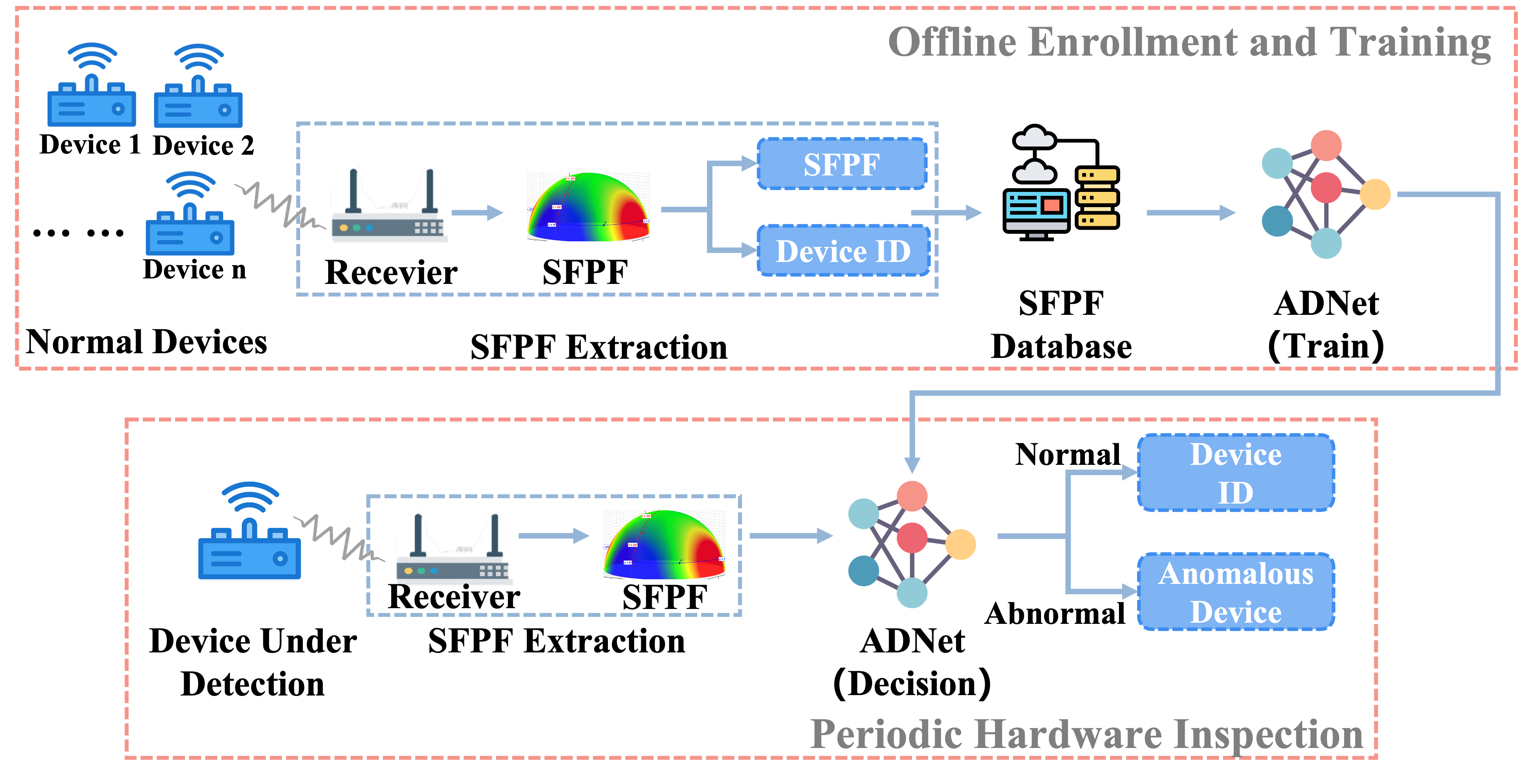}
 \caption{Offline enrollment and training with periodic hardware inspection.}
 \label{fig:system}
\end{figure}

\subsection{Detection Framework}
\label{subsec:detection_framework}

Figure~\ref{fig:system} illustrates the overall SFPF-based hardware anomaly 
detection framework, which consists of offline enrollment and training followed by periodic hardware inspection. 
During offline enrollment and training, SFPFs are collected only from authorized devices 
under their legitimate hardware configurations. Each SFPF is associated with 
the corresponding device identity and used to train the detector to learn the 
hardware-dependent characteristics of enrolled devices.

During periodic inspection, the device under test is measured under the same acquisition settings, and its SFPF is constructed from the received signals. The trained detector then determines whether 
the observed SFPF is consistent with an enrolled device. If so, the 
corresponding device identity is reported; otherwise, the device is rejected 
as anomalous.

\subsection{Detection Network}
\label{subsec:detection_network}

The anomalous-device detection network (ADNet) is a coordinate-aware set network for multi-directional SFPF. Its shared encoder and symmetric pooling follow the permutation-invariant set formulation in~\cite{zaheer2017deepsets}, while coordinate fusion retains the physical direction of each observation. Concatenating the real and imaginary parts of \(\widetilde{\mathbf S}\) along the sample dimension gives
\(\bX\in\mathbb R^{N_a\times N_f\times 2N_s}\). Each directional observation
\(\bX_a\in\mathbb R^{N_f\times 2N_s}\) is processed independently by a shared three-layer 1-D CNN:
\[
N_f\xrightarrow[\;s=4\;]{k=9}32
\xrightarrow[\;s=4\;]{k=7}64
\xrightarrow[\;s=4\;]{k=5}128.
\]
Each convolution is followed by batch normalization and GELU activation. Adaptive average pooling then produces a 128-dimensional directional feature
\begin{equation}
 \mathbf e_a=G_{\mathrm{cnn}}(\bX_a)\in\mathbb R^{128}.
 \label{eq:direction_encoding}
\end{equation}

To retain the physical observation direction, each feature is augmented with its unit direction coordinate
\begin{equation}
 \mathbf c_a=
 [\sin\theta_a\cos\phi_a,\,
  \sin\theta_a\sin\phi_a,\,
  \cos\theta_a]^{\mathsf T}.
 \label{eq:direction_coordinate}
\end{equation}
The directional feature and coordinate are fused through a shared linear projection:
\begin{equation}
 \mathbf q_a=
 \operatorname{GELU}\!\left(
 \operatorname{LN}\!\left(
 W_c[\mathbf e_a;\mathbf c_a]+\mathbf b_c
 \right)\right)\in\mathbb R^{128}.
 \label{eq:coordinate_fusion}
\end{equation}

The directional feature set is summarized using complementary mean and maximum pooling,
\begin{equation}
 \mathbf h=\psi\!\left(
 \left[
 \frac{1}{N_a}\sum_{a=1}^{N_a}\mathbf q_a\,;\,
 \max_{1\leq a\leq N_a}\mathbf q_a
 \right]\right)\in\mathbb R^{128},
 \label{eq:set_pooling}
\end{equation}
where \(\psi(\cdot)\) is a \(256\)-to-\(128\) projection followed by layer normalization, GELU, and dropout. Mean pooling represents the overall device response, whereas maximum pooling retains localized directional variations caused by hardware replacement.

A linear classifier maps \(\mathbf h\) to the \(K\) enrolled device classes. The network is trained using only enrolled devices and the cross-entropy loss
\begin{equation}
 \mathcal L_{\mathrm{cls}}
 =-\frac{1}{N}\sum_{i=1}^{N}
 \log p_{y_i}(\bX_i).
 \label{eq:classification_loss}
\end{equation}
No anomalous samples or auxiliary self-supervised objectives are required during training.

ADNet rejects anomalous devices using class-conditional Mahalanobis distance~\cite{lee2018mahalanobis} in the learned feature space. Let \(\boldsymbol\mu_k\) denote the feature mean of enrolled class \(k\), and let \(\boldsymbol\Sigma\) be the shared diagonal covariance estimated from known validation samples. The anomaly score is
\begin{equation}
 d(\mathbf h)=
 \min_{1\leq k\leq K}
 (\mathbf h-\boldsymbol\mu_k)^{\mathsf T}
 \boldsymbol\Sigma^{-1}
 (\mathbf h-\boldsymbol\mu_k).
 \label{eq:mahalanobis_score}
\end{equation}
The rejection threshold \(\gamma\) is determined exclusively from known-device validation scores. The final decision is
\begin{equation}
 y^*=
 \begin{cases}
  \arg\max_k p_k(\bX), & d(\mathbf h)\leq\gamma,\\
  \mathrm{anomalous}, & d(\mathbf h)>\gamma.
 \end{cases}
 \label{eq:open_set_decision}
\end{equation}

\begin{figure*}[t]
 \centering
 \includegraphics[width=1\textwidth]{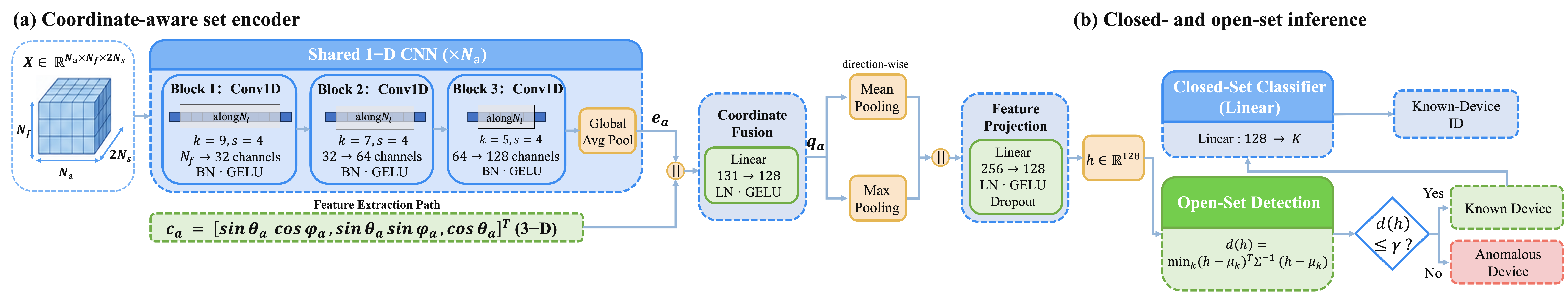}
 \caption{ADNet architecture. A shared 1-D CNN encodes each directional SFPF observation, coordinate fusion retains its physical direction, and mean--maximum pooling forms the device representation. A linear classifier identifies enrolled devices, while Mahalanobis-distance thresholding rejects anomalous devices.}
 \label{fig:adnet}
\end{figure*}

\section{Experimental Evaluation}

\subsection{Simulation Validation of SFPF}

We conduct simulations using Ansys Electronics 2022 R1. To construct PF and SFPF samples, data are collected over multiple observation angles and frequency points. Specifically, $\theta$ ranges from \(0^\circ\) to \(90^\circ\), while $\phi$ ranges from \(0^\circ\) to \(360^\circ\), with a default angular interval of \(1^\circ\). The frequency range is 913--917~MHz with a 0.5-MHz interval. To emulate the attack, a subtle structural modification is introduced into the antenna. The data before and after modification are used to construct the corresponding SFPF samples for the normal and anomalous hardware states, respectively.

\begin{figure*}[t]
 \centering
 \subfloat[Enrolled hardware]{\includegraphics[width=0.38\textwidth]{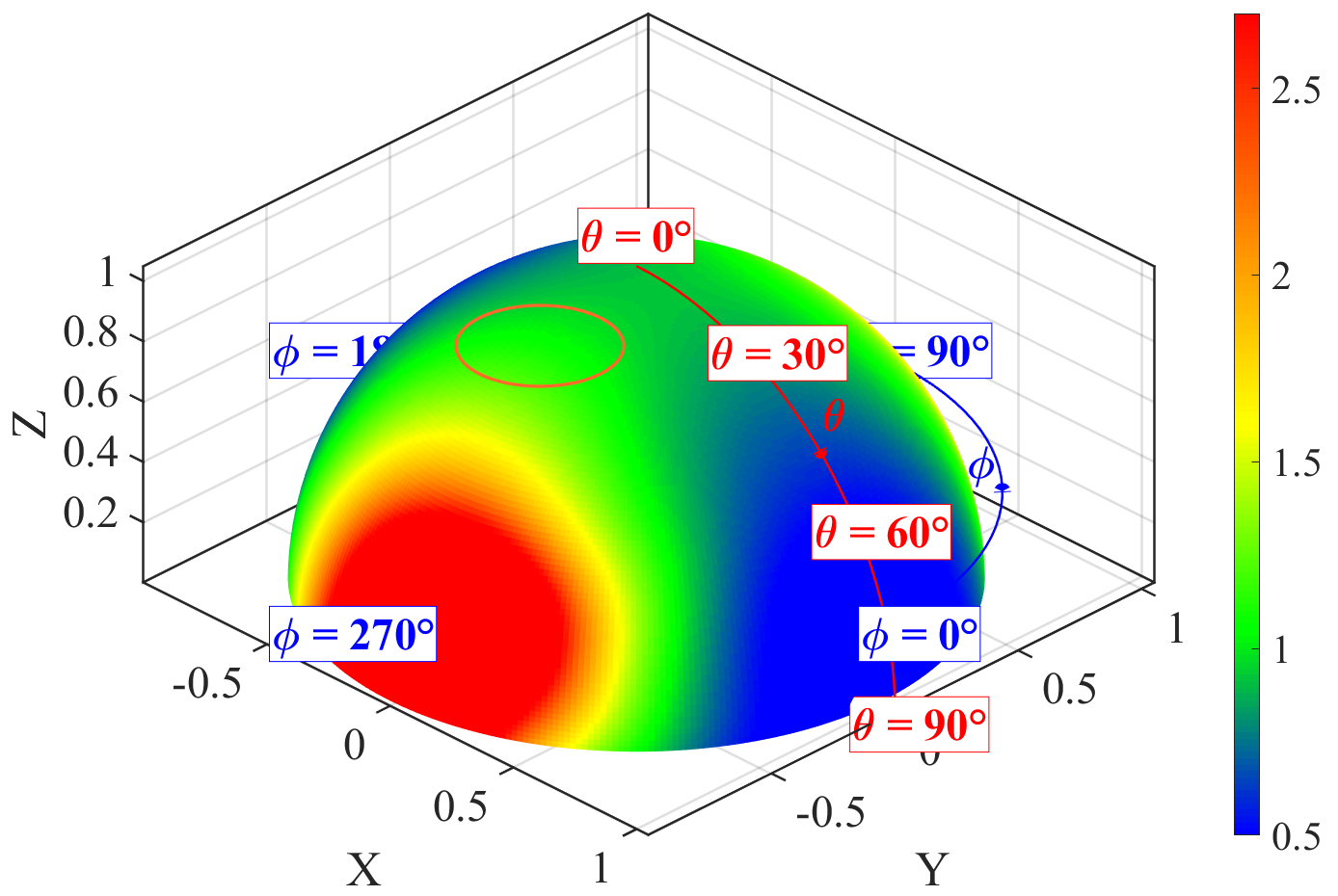}}
 \hfill
 \subfloat[Modified hardware]{\includegraphics[width=0.38\textwidth]{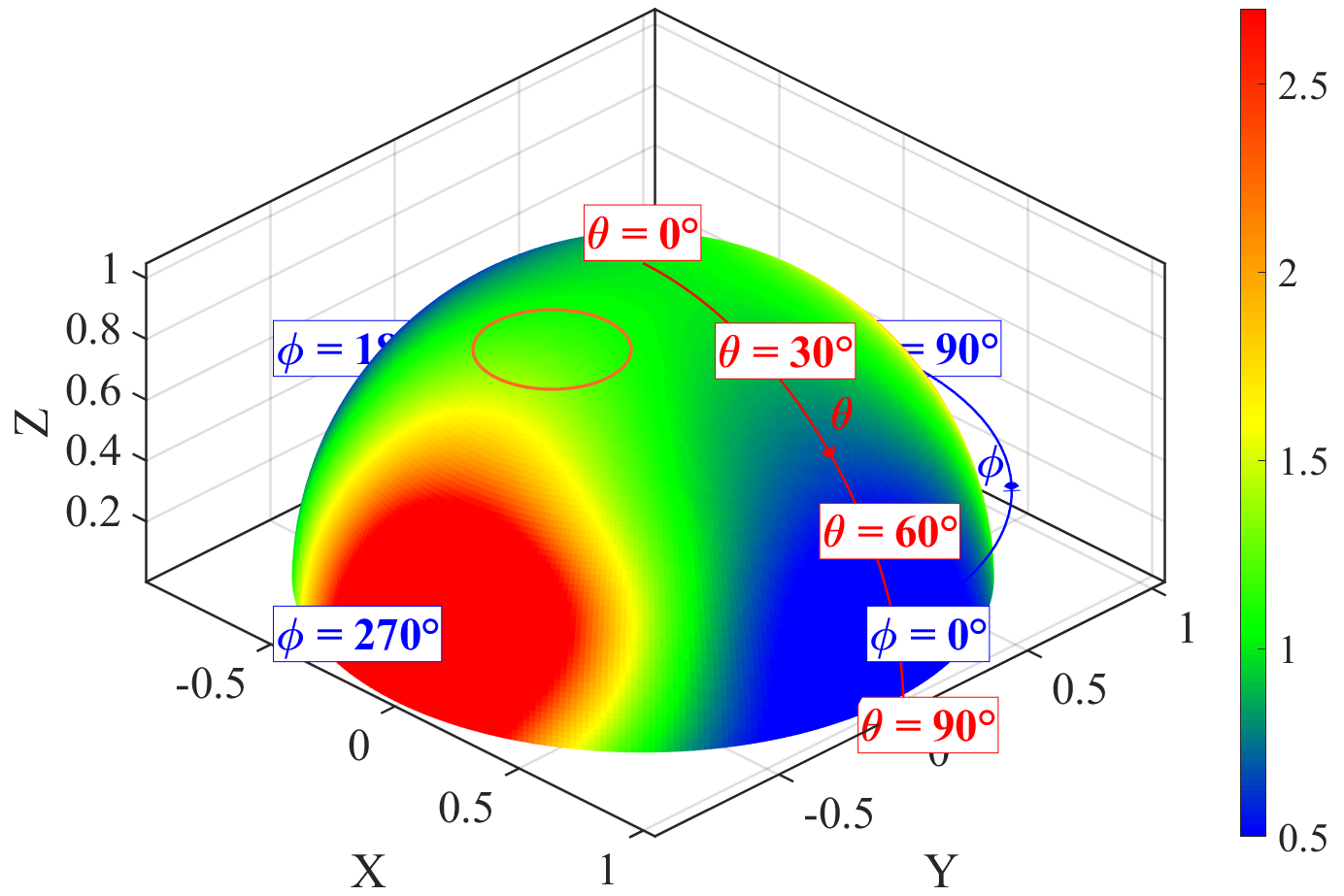}}
 \caption{SFPF distributions before and after hardware perturbation. The highlighted local deviation illustrates direction-dependent hardware sensitivity.}
 \label{fig:sfpf_maps}
\end{figure*}

Figure~\ref{fig:sfpf_maps} shows that the overall response remains similar after modification, while a localized region changes visibly. This behavior agrees with \eqref{eq:sensitivity}: hardware perturbation produces different polarization deviations across the angle--frequency domain, and a fixed-direction PF may miss an informative region.

\begin{figure*}[t]
 \centering
 \subfloat[Multi-angle RFF difference]{\includegraphics[width=0.39\textwidth]{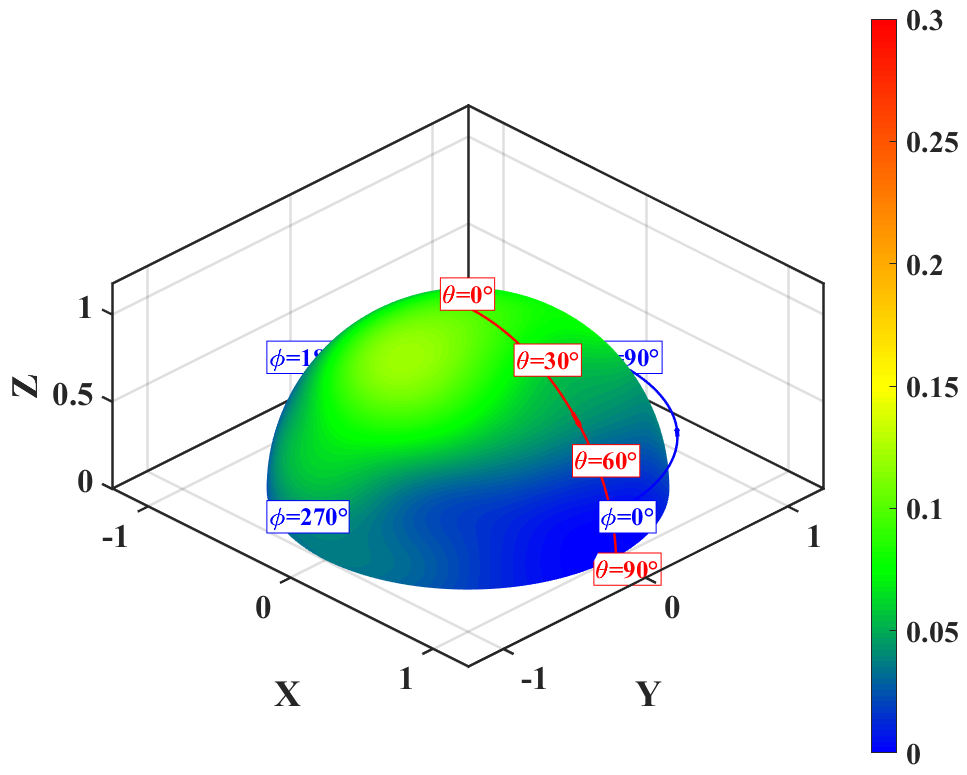}}
 \hfill
 \subfloat[SFPF difference]{\includegraphics[width=0.39\textwidth]{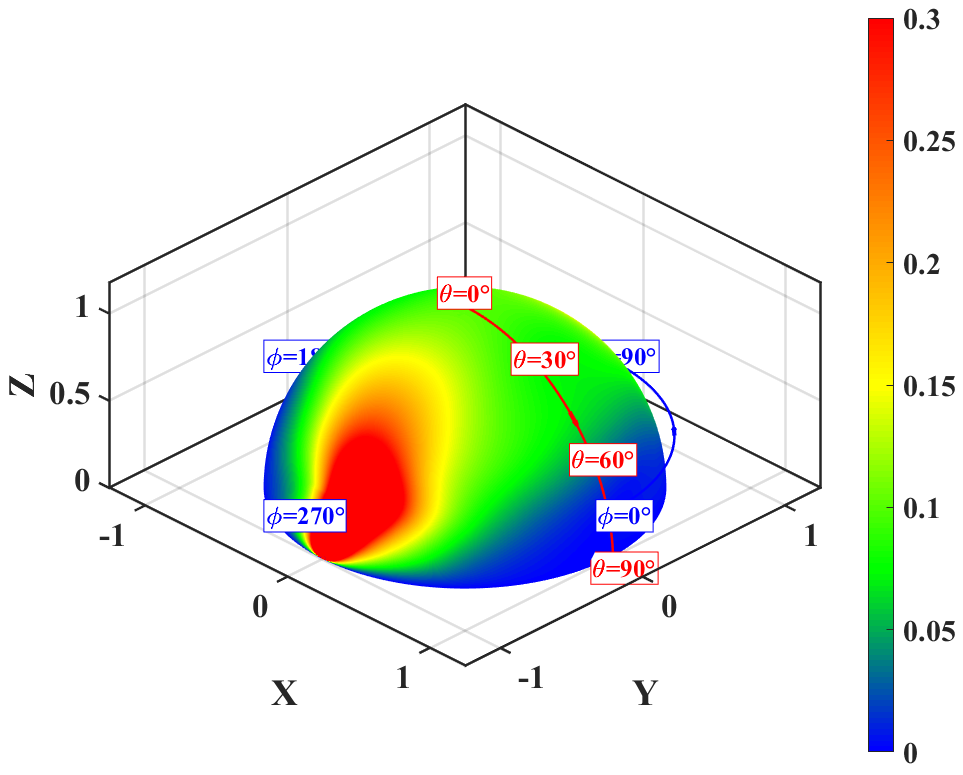}}
 \caption{Normal-to-anomalous device differences over observation direction under the same hardware perturbation. }
 \label{fig:rff_pf_sensitivity}
\end{figure*}

To compare RFF and SFPF under the same hardware perturbation, we evaluate the normal-to-anomalous difference at each observation direction. Figure~\ref{fig:rff_pf_sensitivity} shows that the multi-angle RFF difference is relatively weak over most directions, while SFPF presents larger differences in several spatial regions. The results indicate that SFPF is more sensitive to subtle hardware changes and can reveal variations that are less evident in RFF.

\subsection{Spatio-Frequency Sampling Analysis}

We compare four sampling modes, each using the same fixed budget of nine angle--frequency observations: single-frequency single-angle (SF-SA), single-frequency multi-angle (SF-MA), multi-frequency single-angle (MF-SA, i.e., PF), and multi-frequency multi-angle (MF-MA, i.e., SFPF). SF-SA consists of nine repeated observations acquired at 915~MHz and \((\theta,\phi)=(50^\circ,45^\circ)\); 
SF-MA uses 915~MHz with 
\(\theta\in\{20^\circ,50^\circ,80^\circ\}\) and 
\(\phi\in\{0^\circ,45^\circ,90^\circ\}\); 
MF-SA uses nine frequencies from 913 to 917~MHz at 0.5-MHz intervals at a fixed observation direction; 
and MF-MA uses \(914.5\), \(915\), and \(915.5\)~MHz at 
\((50^\circ,0^\circ)\), \((50^\circ,45^\circ)\), and \((50^\circ,90^\circ)\).

\begin{figure*}[t]
 \centering
 \subfloat[SF-SA]{\includegraphics[width=0.22\textwidth]{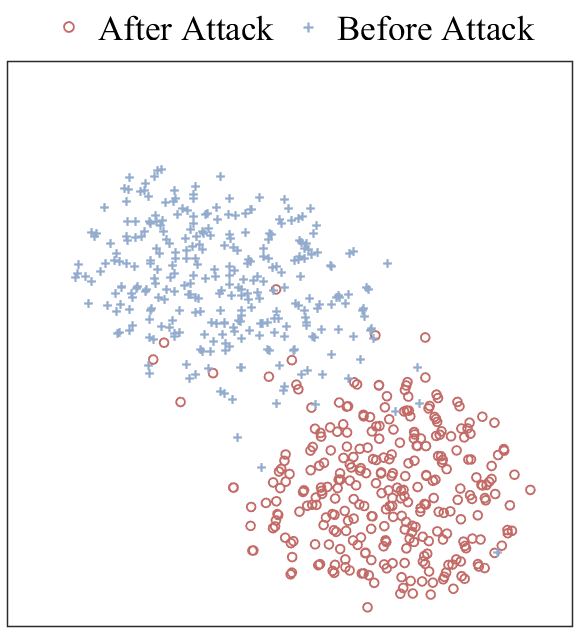}}
 \hfill
 \subfloat[SF-MA]{\includegraphics[width=0.22\textwidth]{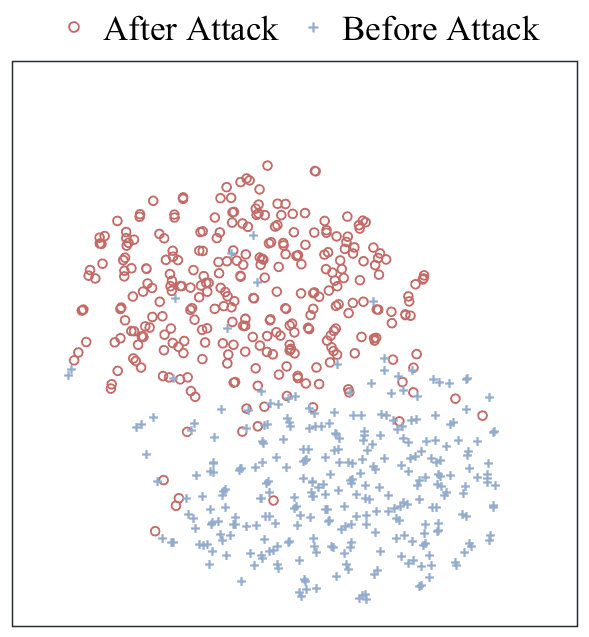}}
 \hfill
 \subfloat[MF-SA (PF)]{\includegraphics[width=0.22\textwidth]{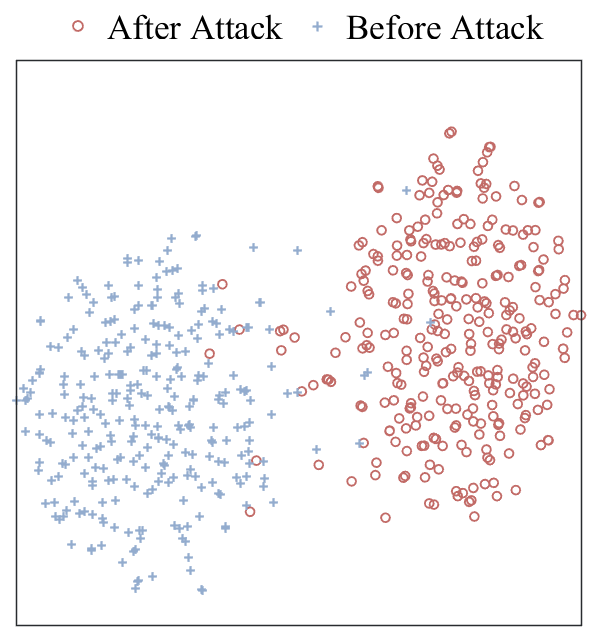}}
 \hfill
 \subfloat[MF-MA (SFPF)]{\includegraphics[width=0.22\textwidth]{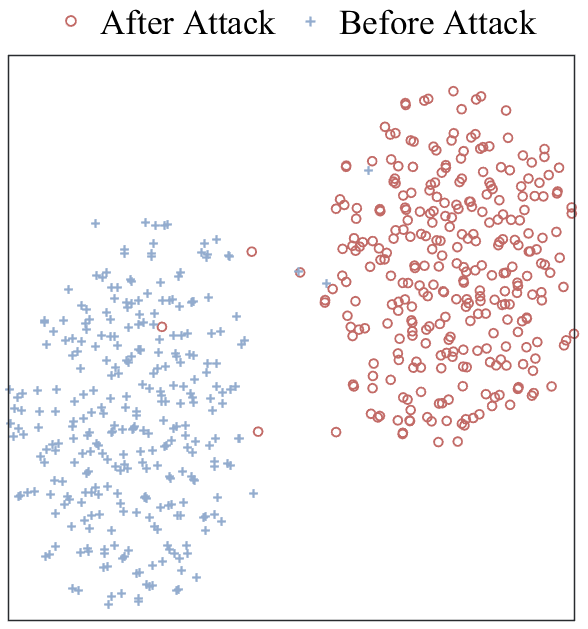}}
 \caption{Feature distributions under four sampling modes.}
 \label{fig:distribution}
\end{figure*}

For pre- and post-modification fingerprints \(\mathbf p_b\) and \(\mathbf p_a\), the energy-normalized complex distance is
\begin{equation}
 d_{\mathrm N}=
 \frac{\|\mathbf p_b-\mathbf p_a\|_2}
 {\sqrt{(\|\mathbf p_b\|_2^2+\|\mathbf p_a\|_2^2)/2}}.
 \label{eq:normalized_distance}
\end{equation}
Table~\ref{tab:separability} shows that MF-MA ranks first on all metrics. Relative to fixed-direction PF, it improves normalized distance by 17.7\%, Fisher score by 45.8\%, and the inter-/intra-class ratio by 11.3\%. These results show that joint frequency--direction sampling improves the separability of SFPF features.

\begin{table}[t]
 \centering
 \caption{Feature separability under different sampling modes.}
 \label{tab:separability}
 \footnotesize
 \setlength{\tabcolsep}{3.0pt}
 \begin{tabular}{lccc}
 \toprule
 Mode & \(d_{\mathrm N}\) & Fisher score & Inter/intra ratio \\
 \midrule
 SF-SA & 0.1070 & 4.3834 & 3.2991 \\
 SF-MA & 0.1203 & 2.3915 & 2.2896 \\
 MF-SA (PF) & 0.1072 & 4.8699 & 3.6527 \\
 MF-MA (SFPF) & \textbf{0.1261} & \textbf{7.1015} & \textbf{4.0654} \\
 \bottomrule
 \end{tabular}
\end{table}

\begin{figure*}[t]
 \centering
 \subfloat[$\theta$ coverage]{\includegraphics[width=0.31\textwidth]{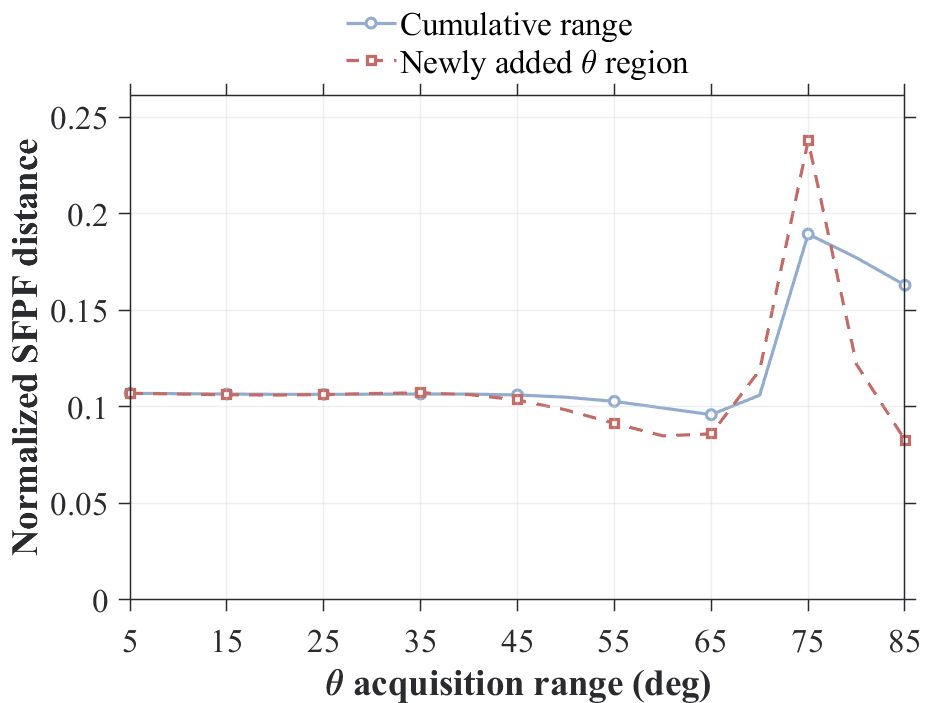}}
 \hfill
 \subfloat[$\phi$ coverage]{\includegraphics[width=0.31\textwidth]{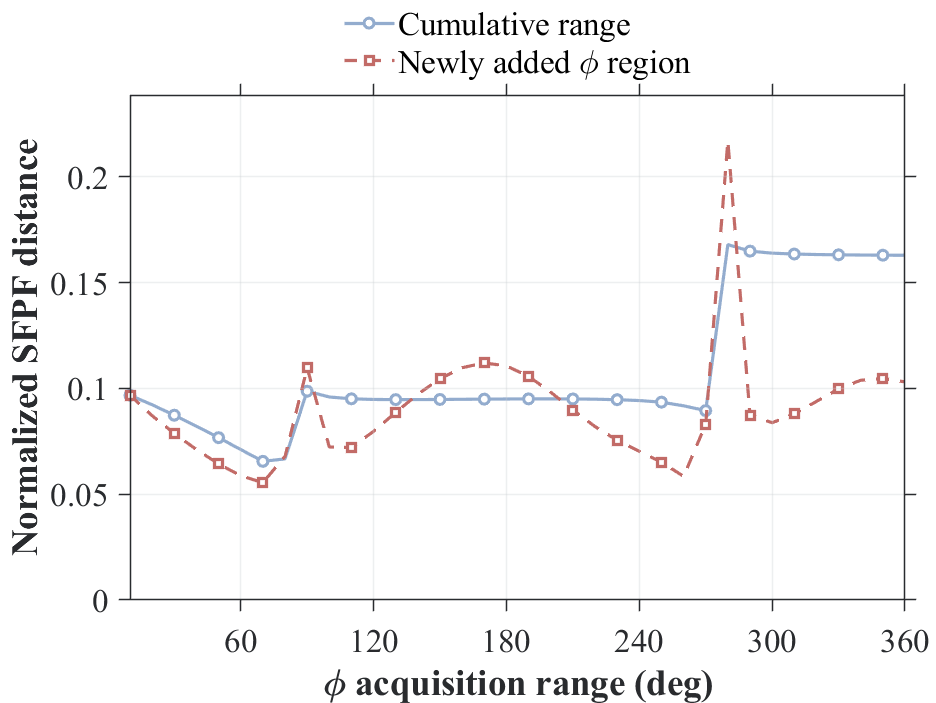}}
 \hfill
 \subfloat[Angular interval]{\includegraphics[width=0.31\textwidth]{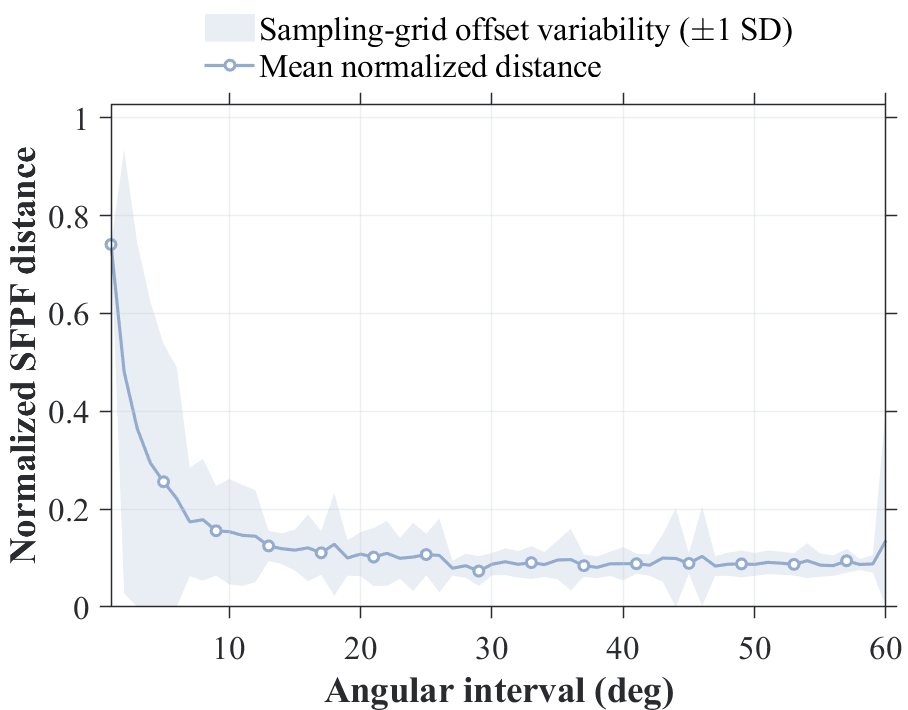}}
 \caption{Effect of spatial acquisition parameters on normalized SFPF difference.}
 \label{fig:spatial_parameters}
\end{figure*}

Figure~\ref{fig:spatial_parameters} further examines the spatial acquisition parameters. Frequency is fixed to 913--917~MHz at 0.5-MHz intervals. Expanding $\theta$ or $\phi$ does not improve separability uniformly; instead, the distance rises sharply when newly included regions contain sensitive directions. Fine angular sampling better retains these localized differences, whereas coarse grids may skip them. 

\begin{figure*}[t]
 \centering
 \subfloat[Frequency coverage]{\includegraphics[width=0.40\textwidth]{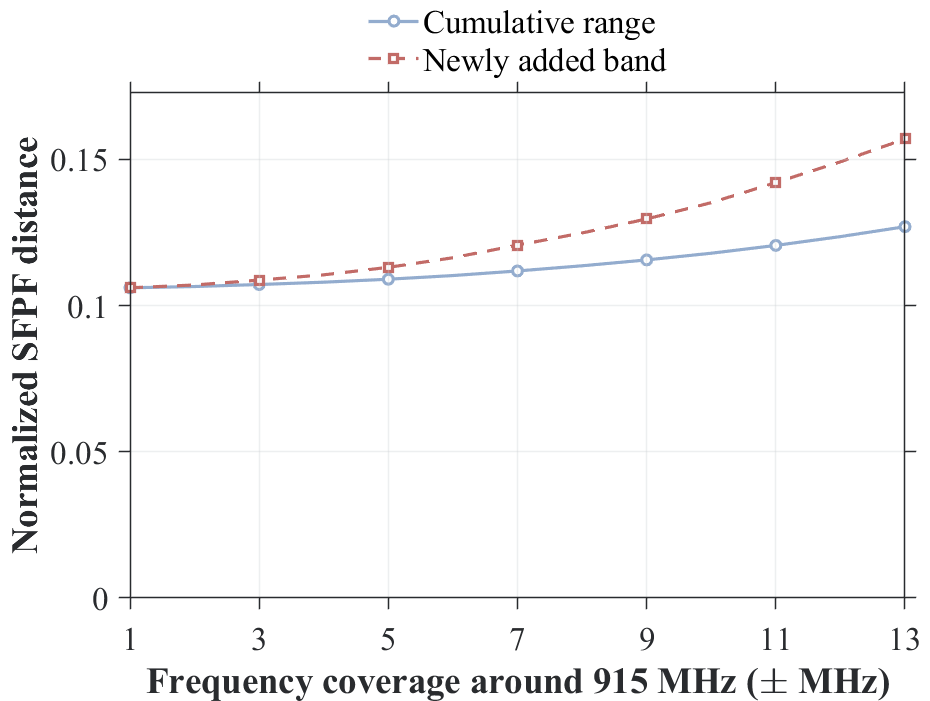}}
 \hfill
 \subfloat[Frequency interval]{\includegraphics[width=0.40\textwidth]{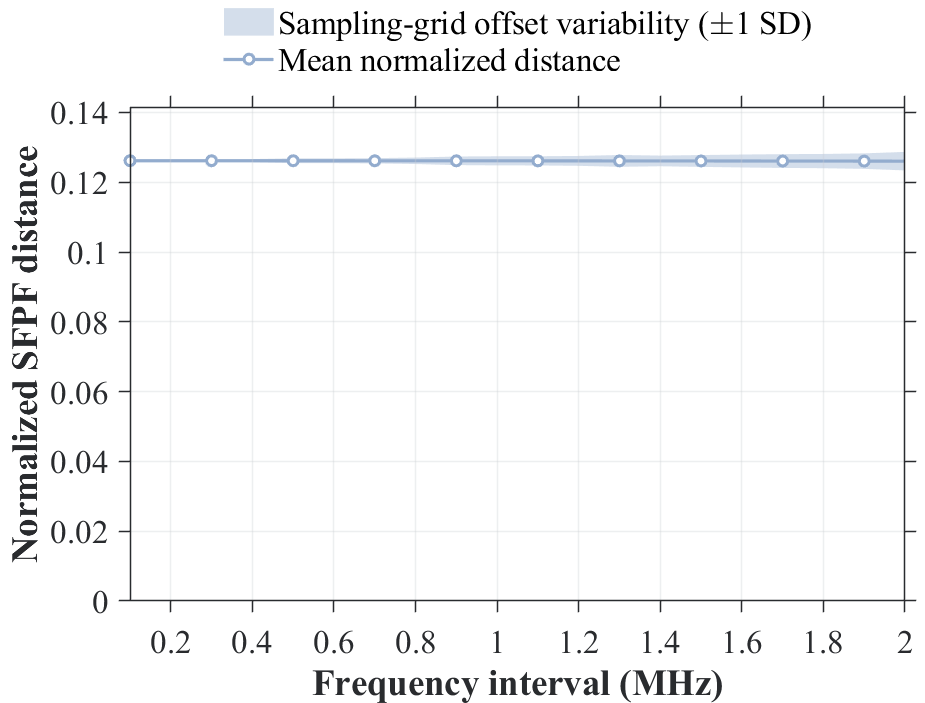}}
 \caption{Effect of frequency acquisition parameters on normalized SFPF difference.}
 \label{fig:frequency_parameters}
\end{figure*}

Figure~\ref{fig:frequency_parameters} provides the frequency domain analysis, with angle fixed to \(\theta\in[0^\circ,80^\circ]\), \(\phi\in[0^\circ,90^\circ]\), and \(5^\circ\) intervals. Wider frequency coverage increases the cumulative distance because it captures more of the modal variation in \eqref{eq:modal_coeff}, and the newly added outer bands are particularly sensitive. By contrast, changing the frequency interval within the same response region leaves the mean distance almost unchanged. These results indicate that covering informative spectral regions is more important than simply increasing sampling density.

\subsection{Experimental Setup}

The prototype contains ten communication devices enrolled with their original hardware. A USRP X310 with an orthogonal dual-polarized antenna is fixed 3~m from the device under test, which is mounted on a motorized pan--tilt platform. Measurements use nine frequencies from 913 to 917~MHz at 0.5-MHz intervals and 1024 complex samples per angle--frequency pair. Elevation covers \(0^\circ\)--\(60^\circ\) and azimuth covers \(0^\circ\)--\(120^\circ\), both at \(5^\circ\) intervals. Merging duplicate pole coordinates gives 301 physical directions per SFPF.

Five devices are selected as attack targets. Each uses its own replacement antenna, RF front end, and digital/baseband module; no module is reused across targets. We evaluate A, R, D, A+R, A+D, R+D, and A+R+D. Only normal samples at 20~dB are used for training, while non-overlapping normal and anomalous samples over 0--20~dB are used for testing.

The coordinate-aware set network contains approximately \(1.1\times10^5\) trainable parameters and is trained for 30 epochs using AdamW with an initial learning rate of \(10^{-3}\), weight decay \(10^{-4}\), and an effective batch size of 64. The learning rate is linearly warmed up for two epochs and then cosine-decayed to \(10^{-5}\); dropout is set to \(0.1\), and the checkpoint with the highest known-device validation accuracy is used for testing. Class means, the shared diagonal covariance, and the rejection threshold are estimated exclusively from known-device validation features, with the threshold set to their 99th-percentile Mahalanobis score.

RFF and SFPF are collected over the same set of observation angles. PF is independently measured at boresight \((0^\circ,0^\circ)\), where both polarization components are reliable, over the same frequency set. PF measurements are repeated at the same observation direction to obtain the same number of samples as SFPF, without introducing additional spatial information. RFF, PF, and SFPF use the same devices for training and testing and the same detection configuration.

\subsection{Hardware Anomaly Detection Results}

\begin{figure*}[t]
 \centering
 \subfloat[Fingerprint comparison]{\includegraphics[width=0.40\textwidth]{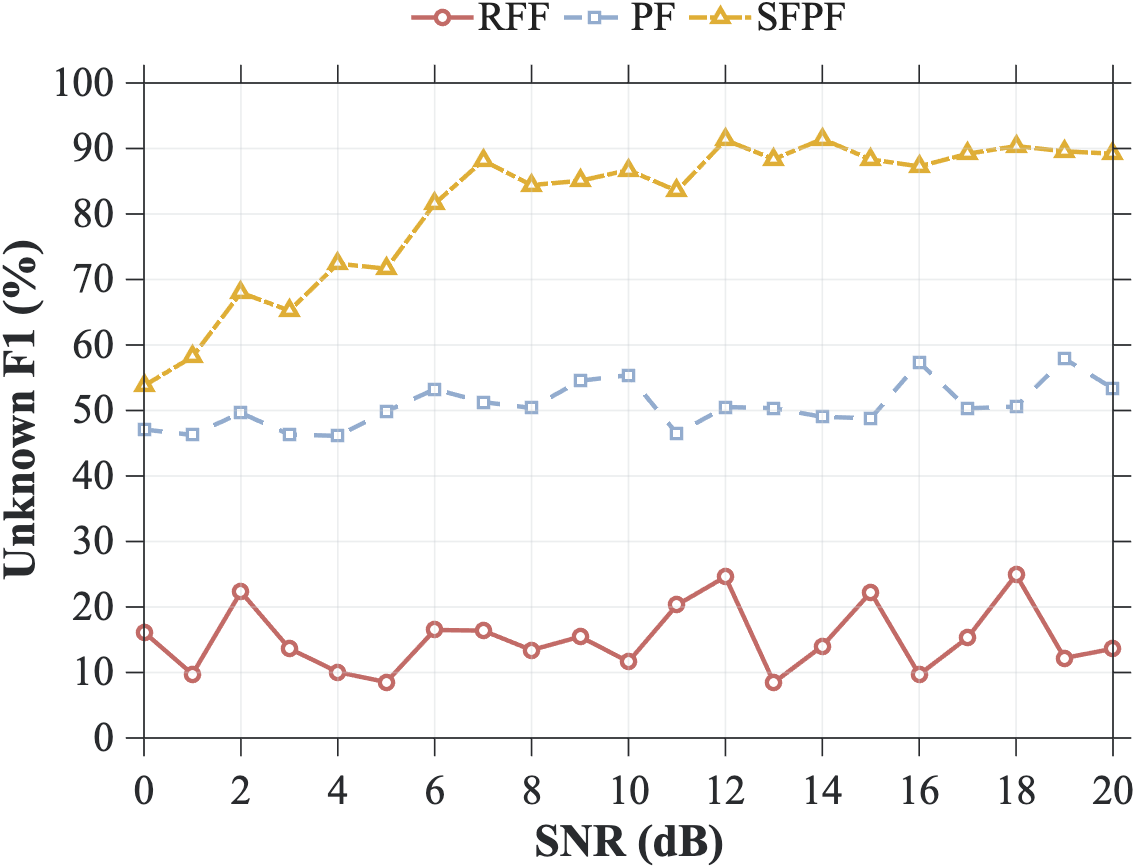}}
 \hfill
 \subfloat[SFPF open-set performance]{\includegraphics[width=0.40\textwidth]{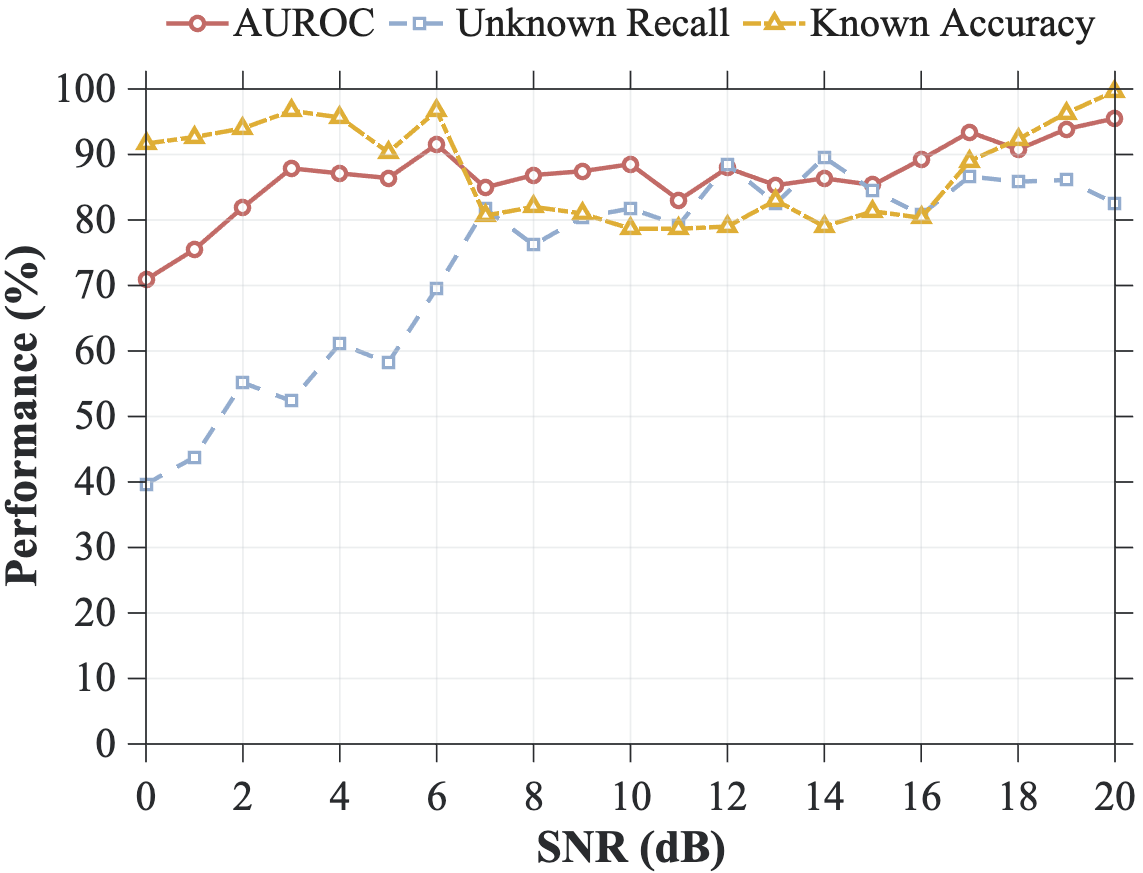}}
 \caption{Detection results over 0--20~dB.}
 \label{fig:fingerprint_results}
\end{figure*}

\begin{figure*}[t]
 \centering
 \subfloat[Detection error rate]{\includegraphics[width=0.31\textwidth]{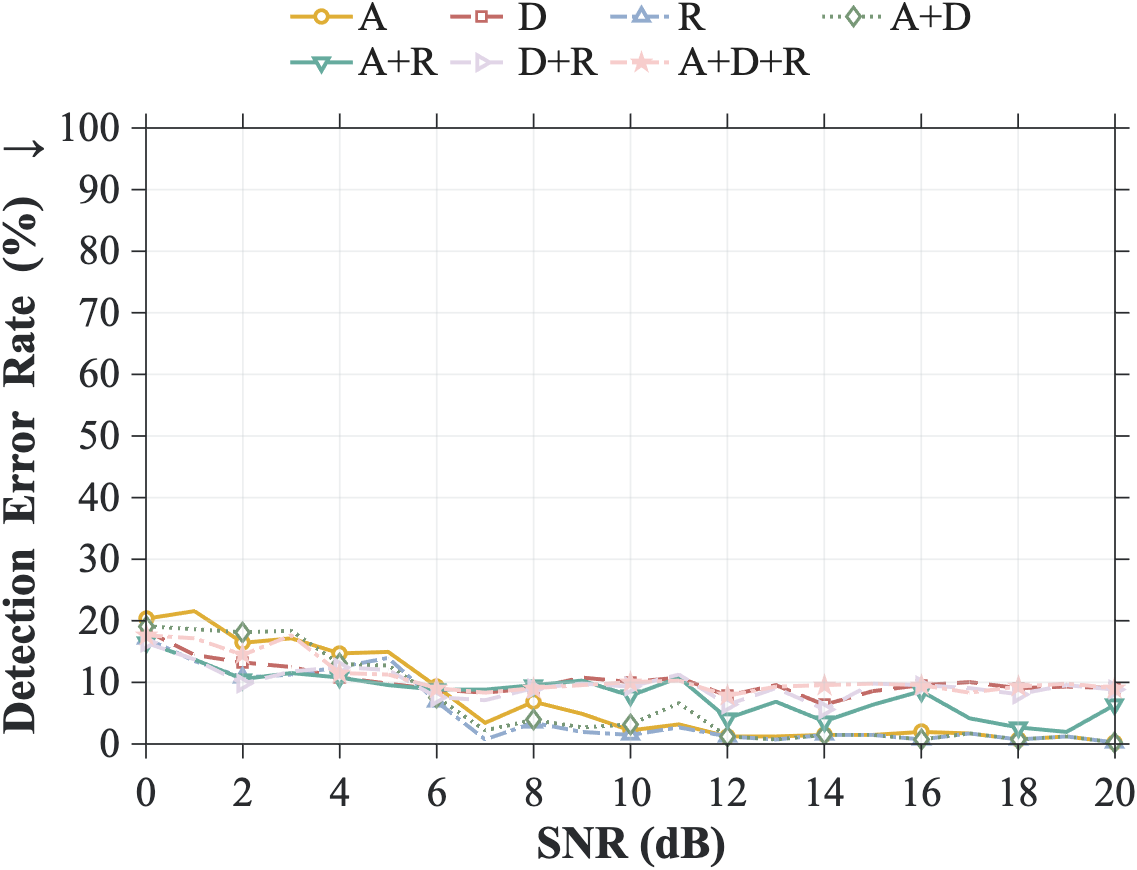}}
 \hfill
 \subfloat[Known-device false alarm rate]{\includegraphics[width=0.31\textwidth]{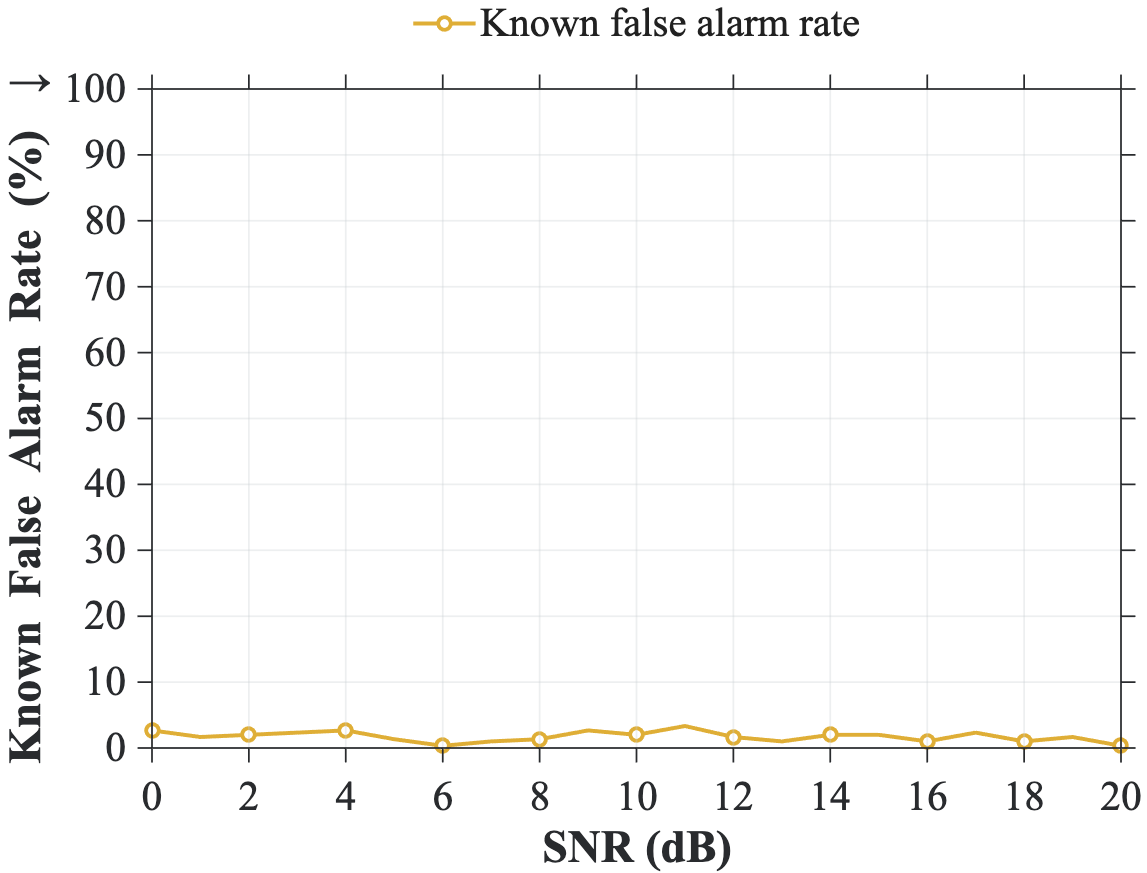}}
 \hfill
 \subfloat[Unknown-device miss rate]{\includegraphics[width=0.31\textwidth]{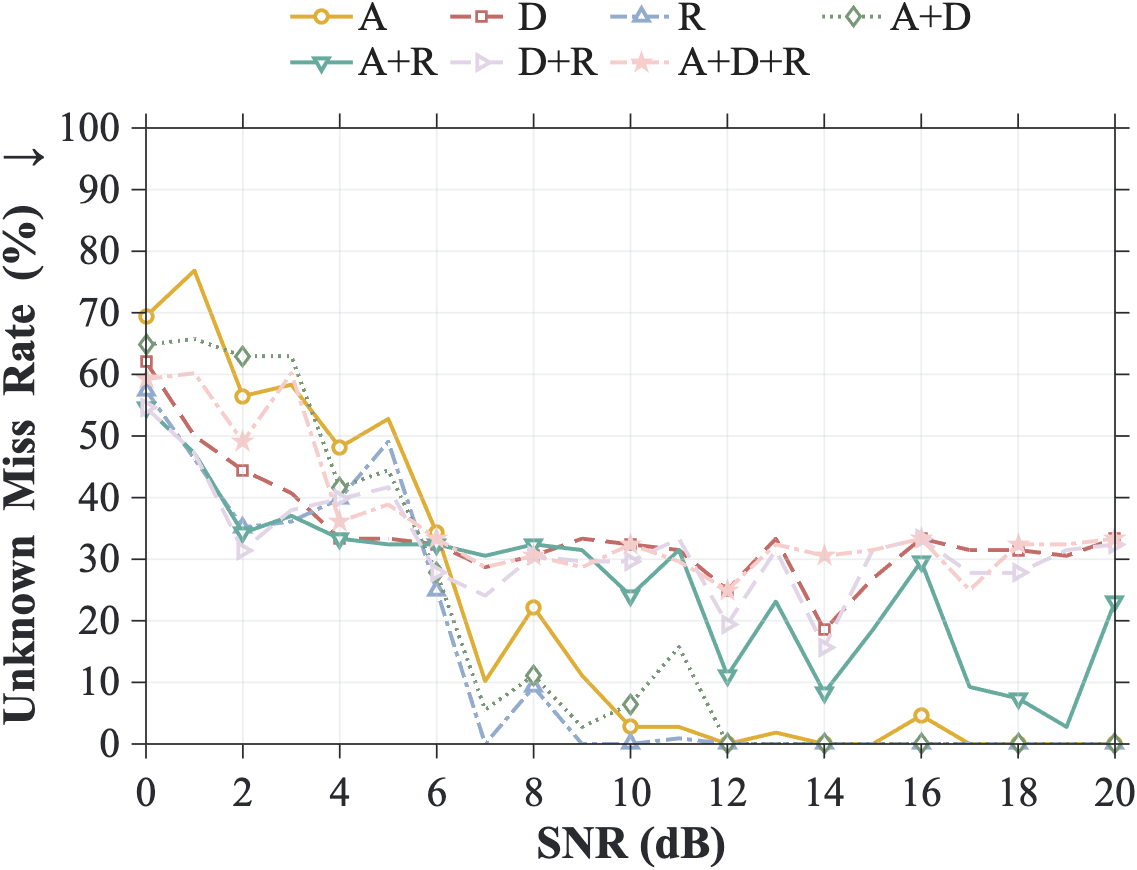}}
 \caption{Security performance under seven hardware-replacement scenarios; lower values are better.}
 \label{fig:security_performance}
\end{figure*}

Figure~\ref{fig:fingerprint_results}(a) compares the anomalous-device F1 scores of RFF, PF, and SFPF. SFPF consistently achieves higher F1 scores across the tested SNR range. At 15--20~dB, the F1 scores of RFF and PF range from 9.7--24.9\% and 48.8--57.9\%, respectively, whereas SFPF achieves 87.3--90.4\%. These results demonstrate that SFPF is more effective in detecting subtle hardware changes.

Figure~\ref{fig:fingerprint_results}(b) presents the open-set detection performance of SFPF. The AUROC increases from 70.9\% at 0~dB to 95.5\% at 20~dB, and ranges from 85.4\% to 95.5\% at 15--20~dB. Over the same SNR range, unknown-device recall ranges from 80.8\% to 86.6\%, while known-device accuracy ranges from 80.3\% to 99.7\%. These results show that SFPF maintains effective known-device identification and anomalous-device detection under different SNR conditions.

\subsection{Security Performance}
To further evaluate the system from a security perspective, 
Fig.~\ref{fig:security_performance} reports the detection error rate 
$\mathrm{DER}=(FP_K+FN_U)/N$, known-device false alarm rate 
$\mathrm{FAR}_K=FP_K/(FP_K+TN_K)$, and unknown-device miss rate 
$\mathrm{MR}_U=FN_U/(TP_U+FN_U)$, where 
$N=TP_U+FN_U+FP_K+TN_K$. At 10--20~dB, the scenario-wise DER ranges 
from 0.25\% to 11.27\%, while the known-device false alarm rate remains 
between 0.33\% and 3.33\% over the entire SNR range. At 15--20~dB, 
the unknown-device miss rate ranges from 0 to 33.33\% across the seven 
replacement scenarios.
\subsection{Discussion}

At 0.05~s per angle--frequency observation, the 301-direction, nine-frequency grid requires 135.45~s for one complete SFPF acquisition. The current acquisition process is intended for periodic hardware inspection, and reducing the acquisition time remains an important direction for future work. The sampling analysis shows that discriminative changes concentrate in selected angle--frequency regions, motivating sparse acquisition in future work.

\section{Conclusion}
This paper presented SFPF for non-intrusive periodic hardware-anomaly inspection of wireless devices. SFPF extends conventional PF from one observation direction to a joint frequency--direction representation. Its formation model and sensitivity analysis show that hardware-dependent modal excitation and far-field projection cause the same hardware change to produce nonuniform polarization differences over frequency and direction. Simulations confirmed this behavior and showed that, SFPF improved normalized distance, Fisher score, and the inter-/intra-class ratio over PF by 17.7\%, 45.8\%, and 11.3\%, respectively. In prototype experiments with ten devices and seven hardware-replacement scenarios, SFPF consistently outperformed RFF and PF; at 15--20~dB, its anomalous-device F1 reached 87.3--90.4\% and its AUROC reached 85.4--95.5\%. These results support SFPF for periodic non-intrusive inspection. The current full-grid acquisition takes 135.45~s; future work will use sensitivity-guided sparse sampling to reduce this time.

\bibliographystyle{IEEEtran}
\bibliography{ref}

\begin{thebibliography}{10}
\providecommand{\url}[1]{#1}
\csname url@samestyle\endcsname
\providecommand{\newblock}{\relax}
\providecommand{\bibinfo}[2]{#2}
\providecommand{\BIBentrySTDinterwordspacing}{\spaceskip=0pt\relax}
\providecommand{\BIBentryALTinterwordstretchfactor}{4}
\providecommand{\BIBentryALTinterwordspacing}{\spaceskip=\fontdimen2\font plus
\BIBentryALTinterwordstretchfactor\fontdimen3\font minus \fontdimen4\font\relax}
\providecommand{\BIBforeignlanguage}[2]{{%
\expandafter\ifx\csname l@#1\endcsname\relax
\typeout{** WARNING: IEEEtran.bst: No hyphenation pattern has been}%
\typeout{** loaded for the language `#1'. Using the pattern for}%
\typeout{** the default language instead.}%
\else
\language=\csname l@#1\endcsname
\fi
#2}}
\providecommand{\BIBdecl}{\relax}
\BIBdecl

\bibitem{withsecure2020fakecisco}
\BIBentryALTinterwordspacing
{F-Secure Consulting}, ``The fake cisco: Hunting for backdoors in counterfeit cisco devices,'' F-Secure Consulting, Tech. Rep., 2020, accessed: 2026-06-03. [Online]. Available: \url{https://labs.withsecure.com/content/dam/labs/docs/2020-07-the-fake-cisco.pdf}
\BIBentrySTDinterwordspacing

\bibitem{tehranipoor2010survey}
M.~Tehranipoor and F.~Koushanfar, ``A survey of hardware trojan taxonomy and detection,'' \emph{IEEE Design \& Test of Computers}, vol.~27, no.~1, pp. 10--25, 2010.

\bibitem{zhou2021backside}
B.~Zhou, A.~Aksoylar, K.~Vigil, R.~Adato, J.~Tan, B.~Goldberg, M.~S. {\"U}nl{\"u}, and A.~Joshi, ``Hardware trojan detection using backside optical imaging,'' \emph{IEEE Transactions on Computer-Aided Design of Integrated Circuits and Systems}, vol.~40, no.~1, pp. 24--37, 2021.

\bibitem{soltanieh2020review}
N.~Soltanieh, Y.~Norouzi, Y.~Yang, and N.~C. Karmakar, ``A review of radio frequency fingerprinting techniques,'' \emph{IEEE Journal of Radio Frequency Identification}, vol.~4, no.~3, pp. 222--233, 2020.

\bibitem{peng2023supervised}
Y.~Peng, C.~Hou, Y.~Zhang, Y.~Lin, G.~Gui, H.~Gacanin, S.~Mao, and F.~Adachi, ``Supervised contrastive learning for rff identification with limited samples,'' \emph{IEEE Internet of Things Journal}, vol.~10, no.~19, pp. 17\,293--17\,306, 2023.

\bibitem{nhem2025explainable}
T.~Nhem, M.~Weyn, M.~Peeters, and R.~Berkvens, ``Explainable rff: Radio frequency fingerprint via spectrogram analysis,'' in \emph{2025 IEEE 36th International Symposium on Personal, Indoor and Mobile Radio Communications (PIMRC)}.\hskip 1em plus 0.5em minus 0.4em\relax IEEE, 2025, pp. 1--6.

\bibitem{wang2025device}
J.~Wang, J.~Xu, D.~Wei, and W.~Huang, ``Device identification based on artificial polarization fingerprint injection,'' in \emph{2025 IEEE Wireless Communications and Networking Conference (WCNC)}.\hskip 1em plus 0.5em minus 0.4em\relax IEEE, 2025, pp. 1--7.

\bibitem{xu2022polarization}
J.~Xu, D.~Wei, and W.~Huang, ``Polarization fingerprint: A novel physical-layer authentication in wireless iot,'' in \emph{2022 IEEE 23rd International Symposium on a World of Wireless, Mobile and Multimedia Networks (WoWMoM)}.\hskip 1em plus 0.5em minus 0.4em\relax IEEE, 2022, pp. 434--443.

\bibitem{xu2022specific}
------, ``Specific emitter identification via spatial characteristic of polarization fingerprint,'' in \emph{2022 IEEE Symposium on Computers and Communications (ISCC)}.\hskip 1em plus 0.5em minus 0.4em\relax IEEE, 2022, pp. 1--7.

\bibitem{agrawal2007trojan}
D.~Agrawal, S.~Baktir, D.~Karakoyunlu, P.~Rohatgi, and B.~Sunar, ``Trojan detection using {IC} fingerprinting,'' in \emph{2007 IEEE Symposium on Security and Privacy}, 2007, pp. 296--310.

\bibitem{huang2014electromagnetic}
H.~Huang, A.~Boyer, and S.~Ben~Dhia, ``The detection of counterfeit integrated circuit by the use of electromagnetic fingerprint,'' in \emph{2014 International Symposium on Electromagnetic Compatibility}, 2014, pp. 1118--1122.

\bibitem{lee2024robust}
D.~Lee, J.~Lee, Y.~Jung, J.~Kauh, and T.~Song, ``Robust hardware trojan detection method by unsupervised learning of electromagnetic signals,'' \emph{IEEE Transactions on Very Large Scale Integration (VLSI) Systems}, vol.~32, no.~12, pp. 2327--2340, 2024.

\bibitem{xu2023polarization}
J.~Xu and D.~Wei, ``Polarization fingerprint-based lorawan physical layer authentication,'' \emph{IEEE Transactions on Information Forensics and Security}, vol.~18, pp. 4593--4608, 2023.

\bibitem{harrington1971theory}
R.~F. Harrington and J.~R. Mautz, ``Theory of characteristic modes for conducting bodies,'' \emph{IEEE Transactions on Antennas and Propagation}, vol.~19, no.~5, pp. 622--628, 1971.

\bibitem{chen2015characteristic}
Y.~Chen and C.-F. Wang, \emph{Characteristic Modes: Theory and Applications in Antenna Engineering}.\hskip 1em plus 0.5em minus 0.4em\relax Hoboken, NJ, USA: John Wiley \& Sons, 2015.

\bibitem{balanis2016antenna}
C.~A. Balanis, \emph{Antenna Theory: Analysis and Design}, 4th~ed.\hskip 1em plus 0.5em minus 0.4em\relax Hoboken, NJ, USA: John Wiley \& Sons, 2016.

\bibitem{ludwig1973definition}
A.~C. Ludwig, ``The definition of cross polarization,'' \emph{IEEE Transactions on Antennas and Propagation}, vol.~21, no.~1, pp. 116--119, 1973.

\bibitem{zaheer2017deepsets}
M.~Zaheer, S.~Kottur, S.~Ravanbakhsh, B.~P{\'o}czos, R.~Salakhutdinov, and A.~J. Smola, ``Deep sets,'' in \emph{Advances in Neural Information Processing Systems}, vol.~30, 2017, pp. 3391--3401.

\bibitem{lee2018mahalanobis}
K.~Lee, K.~Lee, H.~Lee, and J.~Shin, ``A simple unified framework for detecting out-of-distribution samples and adversarial attacks,'' in \emph{Advances in Neural Information Processing Systems}, vol.~31, 2018, pp. 7167--7177.

\end{thebibliography}

\end{document}